\documentclass[conference,10pt]{IEEEtran}
\usepackage{graphicx, array}
\usepackage[colorlinks,urlcolor=blue,linkcolor=blue,]{hyperref}
\usepackage{tabularx}

\usepackage[T1]{fontenc}

\usepackage[utf8]{inputenc}
\usepackage[table]{xcolor}
\ifCLASSINFOpdf
\else
\fi

\def\BibTeX{{\rm B\kern-.05em{\sc i\kern-.025em b}\kern-.08em T\kern-.1667em\lower.7ex\hbox{E}\kern-.125emX}}

\begin{document}
\title{Location Data and COVID-19 Contact Tracing:\\How Data Privacy Regulations and Cell Service Providers Work In Tandem}

\def\plainauthor{Anonymous}

 \author{
 {\rm Callie Monroe}\\
 University of Denver
\\ Email: Callie.Monroe@du.edu
 \and
 {\rm Faiza Tazi}\\
 University of Denver
\\  Email: Faiza.tazi@du.edu
 
 \and
 {\rm Sanchari Das}\\
University of Denver
\\ Email: Sanchari.Das@du.edu
 } 

\IEEEoverridecommandlockouts
\makeatletter\def\@IEEEpubidpullup{6.5\baselineskip}\makeatother
\IEEEpubid{\parbox{\columnwidth}{
    Workshop on Usable Security and Privacy (USEC) 2021 \\
    7 May 2021, Auckland, New Zealand \\
    ISBN 1-891562-73-8 \\
    https://dx.doi.org/10.14722/usec.2021.23010 \\
    www.ndss-symposium.org
}
\hspace{\columnsep}\makebox[\columnwidth]{}}

\maketitle
\pagestyle{plain}

\begin{abstract}
Governments, Healthcare, and Private Organizations in the global scale have been using digital tracking to keep COVID-19 outbreaks under control. Although this method could limit pandemic contagion, it raises significant concerns about user privacy. Known as ~\lq\lq Contact Tracing Apps\rq\rq~, these mobile applications are facilitated by Cellphone Service Providers (CSPs), who enable the spatial and temporal real-time user tracking. Accordingly, it might be speculated that CSPs collect information violating the privacy policies such as GDPR, CCPA, and others. To further clarify, we conducted an in-depth analysis comparing privacy legislations with the real-world practices adapted by CSPs. We found that three of the regulations (GDPR, COPPA, and CCPA) analyzed defined mobile location data as private information, and two (T-Mobile US, Boost Mobile) of the five CSPs that were analyzed did not comply with the COPPA regulation. Our results are crucial in view of the threat these violations represent, especially when it comes to children's data. As such proper security and privacy auditing is necessary to curtail such violations. We conclude by providing actionable recommendations to address concerns and provide privacy-preserving monitoring of the COVID-19 spread through the contact tracing applications. 

\end{abstract}

\vspace{5mm}

\section {Introduction}
\label{sec:intro} 
The security and privacy of personal data are paramount for users in today's age of technology. Thus, understanding users' data privacy, especially for the technological devices they use, such as smartwatches, mobile devices, wearable Internet of Things (IoT) devices, and smart tablets, is very critical~\cite{martin2020data,10.1145/2981547,streiff2019overpowered,hadan2019making,gopavaram2019iotmarketplace}. These devices have infiltrated everyone's daily lives, especially when it comes to mobile devices and are used for multiple day-to-day activities~\cite{yang2010determinants}. According to the GSMA Intelligence report in November of 2019, 91\% of United States consumers own a Smartphone device~\cite{hatt_jarich_2019}. These smartphone devices are connected through Cellphone Service Providers (CSPs) who enable the network connections to communicate and provide the internet connectivity to utilize several smartphone features~\cite{petzer2011perceived}. Given the nature of modern inter-connected communication, these CSPs obtain a lot of personal data from 
different types of interactions~\cite{tsai2010location}. Thus, providing transparency, and ensuring user trust in the privacy policies implemented by these CSPs are of the utmost importance~\cite{sung2020zipphone}. The recent class-action lawsuit against CSPs for selling user historic movement records to third parties further proves the criticality of such situation~\cite{case319cv04063}. 

In addition to the existing privacy concerns, another global phenomenon that has shaken everyone 
due to the stay-at-home order is COVID-19~\cite{khan2020spread,das2020change,karmakar2020evaluating,karmakar2021understanding}. COVID-19 and stay-at-home orders remain a global issue for over a year~\cite{jenkins2021portrait}, where governments and private entities on a national and international level have suggested the use of Mobile Location Data (MLD) for COVID-19 contact tracing efforts through mobile applications~\cite{eames2003contact}. COVID-19 contact tracing applications indicate any proximity to COVID-19 affected patients or track the symptoms of the patients~\cite{keeling2020efficacy}. These applications are often self-reported or track the user's location automatically and apply Machine Learning~\cite{lalmuanawma2020applications}, Graph Theory and Modeling~\cite{rorres2018contact}, or Network Theory~\cite{hebert2020beyond} to determine the spread of the disease, given the nature of how the pandemic spreads~\cite{sahu2020covid}. These applications are often used by organizations to indicate the patients' location or those who came in contact with them to detect and quarantine any possibility of the pandemic spread. Although this method could limit contagion, it raises significant concerns about user privacy~\cite{baumgartner2020mind}.

In addition to the user privacy concerns, there are several ethical concerns over the application, data collection, and algorithms of these applications~\cite{baumgartner2020mind,parker2020ethics,cho2020contact}. These concerns are primarily related to the precise user tracking of the mobile location data, specifically individual identification through location profiling, and being tracked by third party organizations~\cite{sung2020zipphone,zeinalipour2020covid}. In such regards, data regulations and privacy legislation come handy for auditing such location access pertaining to correct application~\cite{shukla2020privacy,bengio2021inherent}. Currently, there are few departments in the United States dedicated explicitly to data protection, and those that do exist are at the state level applying only to that state's residents~\cite{bygrave2010privacy}. Among these are the California Consumer Privacy Act (CCPA) and the New York SHIELD Act (Stop Hacks and Improve Electronic Data Security). Additionally, the United States protects children's data rights through the Children's Online Privacy Protection Act of 1998 (COPPA) and the health data rights of insured patients under the Health Insurance Portability and Accountability Act (HIPAA). These acts are then enforced by the Federal Trade Commission, whose mission is to protect consumers and competitors 

Outside of the US, on a global scale, the General Data Protection Regulation (GDPR) of the European Union, implemented in 2018, is now used as a basis for many companies' privacy policies~\cite{mangini2020empirical}. According to the GDPR, MLD is defined as personal data and thus protected under all articles which address the privacy and security of personal data~\cite{gdpr2018art4}. The GDPR applies to any company servicing European citizens and residents; as such, CSPs should be following the regulations provided by the GDPR for European citizens, if not all their users, despite the location of the EU citizens on which they are based. Another national legislation, the Lei Geral de Proteção de Dados Pessoai(LGPD), is enforced in Brazil and for its citizens worldwide~\cite{LGPD_2019}. The LGPD was heavily influenced by the GDPR and enforces similar regulations regarding MLD~\cite{erickson2018comparative}. Based on these regulations set forth by both international, national, and state-level legislation, it is claimed that data collected by CSPs should be ~\lq\lq specified, explicit, and legitimate\rq\rq~, processed with the assurance of security~\cite{gdpr2016art5}, protected appropriately~\cite{gdpr2018art25}, and all recipients of the data must be disclosed~\cite{gdpr2016art13}. Thus, it is crucial to understand what data privacy regulations exist today and how CSPs collect, process, protect, and distribute any MLD with our focus on the contact-tracing mobile applications. 

In the wake of several privacy legislation such as the GDPR, COPPA, and various other laws in the United States and worldwide, we focused our research on exploring the MLD privacy and security access by CSPs during the COVID-19 pandemic. Such location access is reflected in the COVID-19 contact tracing application usage~\cite{9305893,walrave2020ready}. Subsequently, this paper aims to answer the following research questions:
\begin{itemize}
    \item \textit{RQ1: Are CSP's privacy policies compliant with current data protection legislation?}
    \item \textit{RQ2: In what ways are MLDs collected by the CSPs?}
    \item \textit{RQ3: In what ways are MLDs being processed and protected by the CSPs?}
    \item \textit{RQ4: Which entities are provided the user MLD data collected by CSPs? This RQ particularly wants to explore whether third parties are granted access and if yes, then why and how?}
    \item \textit{RQ5: Do the current policies stated by CSPs align with COVID-19 contact tracing efforts?} 
\end{itemize}

In order to do so, we conducted a detailed analysis of six privacy-focused regulations namely: GDPR~\cite{marcut2018analysis}, LGPD~\cite{raposo2019lgpd}, CCPA~\cite{rothstein2019california}, SHIELD~\cite{lance2019erecting}, COPPA~\cite{reyes2018won}, and HIPAA~\cite{assistance2003summary}, as well as privacy policies of the five largest American CSPs: Verizon Wireless, AT\&T Mobility, T-Mobile US, Boost Mobile, and U.S. Cellular. 
After that, we evaluated how the CSPs comply with these regulations, especially when it comes to MLDs. We found that two of the five CSPs did not comply with the COPPA regulation; however, all the CSPs complied with the data subject's right to know the extent of data being collected.

This study seeks to contribute to a foundational understanding of how data protection regulations and CSP privacy policies interact today, by analyzing the laws and policies related to mobile location data. This paper provides the basis of building guidelines for addressing contact tracing efforts through the use of MLD while respecting existing regulations and addressing the users' privacy concerns.

\section {Related Works}
\label{sec:related}
In this section, we briefly discuss previous mobile location data protection related research. We then provide an overview of the prior literary work while reviewing current and proposed privacy frameworks, and detail an overarching privacy frameworks for location data while emphasizing on MLD-specific frameworks. 
Given our research focus, we finally outline current works discussing COVID-19 contact tracing efforts using mobile location data. 

\subsection {Data privacy frameworks}
Data privacy frameworks are often the best way for organizations to determine how to handle sensitive data most appropriately. From data collection to data handling, utilizing a privacy framework is an integral part of constructing a thorough, transparent, and user-focused privacy policy, as well as judging a privacy policy's effectiveness. Along these lines, Liu proposes a framework for location data privacy founded on location anonymization where they looked into several security- and privacy-focused algorithms such as K-anonymity~\cite{liu2007data}. A similar concept was introduced by Beresford and Stajano~\cite{beresford2003location} with a location privacy protecting framework, based on frequently changing pseudonyms that allow users to be anonymous. They detailed the effectiveness of providing noise into the data set to avoid identification of users based on precise location and user data.

Similarly, Hoepman introduced privacy design strategies~\cite{hoepman2014privacy}, this notion used the existing data protection legislation as a starting point to determine 8 privacy design strategies namely, ~\textit{Minimise}, ~\textit{Hide}, ~\textit{Separate}, ~\textit{Aggregate}, ~\textit{Inform}, ~\textit{Control}, ~\textit{Enforce} and \textit{Demonstrate}. The proposed generalized framework is used for both designing a privacy respecting system as well as evaluating the privacy impact of existing systems.
In a different approach, Ahamed et al. proposes the elimination of third party location anonymizers through the use of probabilistic anonymity that is calculated based on historic Wi-Fi Access Point data~\cite{AhamedSheikh2012Anlp}. The purpose being keeping data from being passed between parties unnecessarily, and instead approach the storage of location data only as Wi-Fi access point locations. In theory this will allow data subjects to maintain anonymity from the get-go. On the other hand, Lee et al. suggests using a location privacy preserving mechanism which receives actual location events and outputs observations or manipulations of this event~\cite{LeeChao2013AFoE}, the goal being to maintain data obfuscation even after an adversary has gained access to the data, allowing for anonymity. 

Shaham et al. also introduced a privacy framework that is specific to spatiotemporal trajectory datasets. Dubbed the Machine Learning Anonymization (MLA)~\cite{shaham2020privacy}, this framework uses  machine learning algorithms for clustering the trajectories, 
to preserve the privacy of location data.
While all these frameworks provide novel solutions to creating and upholding anonymous MLD, ultimately we were  inspired from Cavoukian's privacy framework~\cite{cavoukian_2011} the most, since it was designed to evaluate the privacy impacts of IT systems. Furthermore, the proposed principles of this system are pertinent to our study of CSPs privacy policies.


\begin{enumerate}
\item~\textit{Proactive not Reactive; Preventative not Remedial}: This is meant to anticipate and thwart privacy infractions before the tentative occurrences of the privacy breaches.
\item~\textit{Privacy as the Default}: They mention that it is highly effective to have privacy built into the system by default as a critical component.
\item~\textit{Privacy Embedded into Design}: By discussing the design of tools and technologies Cavoukian mentions, ~\lq\lq Privacy by Design is embedded into the design and architecture of IT systems and business practices\rq\rq~.
\item~\textit{Full Functionality - Positive-Sum, not Zero-Sum}: While discussing about the functionality, usability, and privacy trade-offs Cavoukian writes, ~\lq\lq Privacy by Design seeks to accommodate all legitimate interests and objectives in a positive-sum ~\lq\lq winwin\rq\rq~ manner, not through a dated, zero-sum approach, where unnecessary trade-offs are made\rq\rq~
\item~\textit{End-to-End Security - Life-cycle Protection}: In addition to the privacy, it is also important to understand security of these IT systems, tools, and technologies. Though colloquially many might mention that privacy and security are inversely proportional, this is not true. Thus, in this framework, the author details on the development of security measures throughout the life-cycle of the data which the users want protected.
\item~\textit{Visibility and Transparency}: While discussing on some auditing protocols, the author says that organizations should be ~\lq\lq operating according to the stated promises and objectives\rq\rq~. This is also critical from the legal perspectives.
\item~\textit{Respect for User Privacy}: This is extremely critical from multiple vectors, such as the user side, ethical and legal perspective, as well as software development perspective. Thus, Cavoukian discusses, that ~\lq\lq Above all, Privacy by Design requires architects and operators to keep the interests of the individual uppermost by offering such measures as strong privacy defaults\rq\rq~
\end{enumerate}

\subsection {Analysis of location privacy laws}
Since a large portion of this study focuses on the analysis of data privacy legislation, it is important to understand how these regulations have previously been analysed for their implications on MLD. There are few previous studies that look at legislation specifically as it impacts MLD, and the ones that do exist focus exclusively on the GDPR. Georgiadou et al. look at the implementations of the GDPR and proceed to analyse location privacy at the individual and cultural levels through the eyes of an individual data subject "Alice"~\cite{GeorgiadouYola2019LPit}. 
from the perspective of the African Union's relationship with data protection legislation where they propose that data protection rights should be afforded to all people, not just those of the EU or the US. 

In their paper, Reyes et al. analyze mobile applications' compliance with COPPA, including a geo-location analysis of these apps. The paper examines the general conformity of android applications with the COPPA regulation   ~\cite{reyes2018won}. Similarly Apthorpe et al. and Streiff et al. evaluated the compliance of Internet of Things (IoT) toys' privacy norms with COPPA~\cite{apthorpe2019evaluating,streiff2018s}. Finally, Liccardy et al. show the difficulties developers are facing to comply with the general rules of COPPA~\cite{liccardi2014can}. While these papers, review the general rules of COPPA, we were unable to find a paper that specifically targets location data as it relates to the COPPA regulation.  

Ataei et al. examine the GDPR legislation in depth to understand how certain aspects can be implemented using user interfaces(UI), and analyze how best to comply with GDPR regulation~\cite{AtaeiMehrnaz2018CwPL}. Consequently, Ataei et al. defined a set of guidelines for Location Based Service(LBS) design and development, with the goal of making it easier for developers to create systems that are GDPR compliant. These guidelines fall into three categories: notice, consent, and control groups. 


Both Georgiadou et al.~\cite{GeorgiadouYola2019LPit} Ataei et al.~\cite{AtaeiMehrnaz2018CwPL} and studies provide a thorough analysis of GDPR policy as it affects location. Georgiadou et al.~\cite{GeorgiadouYola2019LPit} uses this analysis to encourage privacy legislation developments in the African Union while Ataei et al.~\cite{AtaeiMehrnaz2018CwPL} uses the analysis of the GDPR to help produce usable UI. While the GDPR is a world leading legislation, it limits both studies to a legislation that is enforceable only in the European union and for EU citizens. 

As discussed, the majority of these studies primarily focus on the GDPR which can be limiting when addressing policy issues in countries outside of the EU. While we will take a look at the GDPR legislation, we will also look at four additional pieces of legislation that are enforced in the US as well as a legislation that is enforced in Brazil. The GDPR provides an excellent framework for data privacy regulation, however, it is critical to look at multiple legislation to get a thorough understanding of how data privacy is protected on the state level in the United States as well as the national level across the world. Thus, for our analysis we have looked into legislation which are applied in state-wide, national, and international perspective including SHIELD ACT, CCPA, HIPAA, COPPA, LGPD, and GDPR.

\subsection {COVID-19 contact tracing efforts with MLD}
Since the emergence of COVID-19, controlling and stopping the spread of the virus has been the primary concern of many government officials, medical professionals, and media outlets in the US and across the globe. Among these discussions, a prevalent topic has been the use of mobile location data to help develop accurate contact tracing methods through mobile-based applications. This proposal brings forth privacy concerns both from the perspective of users as well as CSPs in what data is tracked ~\cite{harari2020process}, how the data is collected and analyzed ~\cite{thaichon2016analysis}, and whether there is any third party access involved~\cite{whaiduzzaman2014measuring}. However, since the notion of using MLD to track infection and more specifically COVID-19 cases, is a novel concept, the studies that have been published about this topic are sparse.

Along these concepts, Egan discusses concerns of companies adhering to their stated privacy policies as many companies began to offer up their MLD in aggregate, anonymized forms~\cite{CaseyEgan2020Mldt}. However, Egan states that not all claims of data anonymization are true or sufficient enough to deter all adversaries. Egan presents the FTC's case against Facebook Inc. and Cambridge Analytica LLC as prime example of data misuse despite policy claims~\cite{CaseyEgan2020Mldt}. Egan also addresses how a private company in New York, Unacast, has already began using ~\lq\lq anonymous device location data\rq\rq~ to develop a ~\lq\lq Social Distancing Scoreboard\rq\rq~ in order to assess social distancing within a certain region~\cite{CaseyEgan2020Mldt}. Unacast claims that all MLD being used is anonymous and comes in an aggregate form~\cite{unacast}. 

On the other hand, Oliver et al. provide a more thorough look at how MLD can help create preventative measures during COVID-19, why it hasn't seen widespread implementation, and how widespread implementation can become possible~\cite{OliverNuria2020Mpdf}.They determine that MLD is best used to understand which individuals have been infected and who they came in contact with, how effective implementing mobility and social restrictions are, and how lifting restrictions affect behavior. However, despite these advantageous uses, there remains the issues of overwhelming demand for government officials to make these critical decisions, lack of data access through CSPs, and public concerns on data privacy, protection, and the civil liberties of the public~\cite{OliverNuria2020Mpdf}. Finally, Oliver et al. suggest the use of mixed teams of government officials, CSPs, and technology companies to tackle the issue of MLD for COVID-19 preventative measures.

Oliver et al. and Egan provide thorough overviews of the implementations and concerns of using MLD for COVID-19 contact tracing, despite this being a very new field. Our study furthers the contribution to this developing field by determining if COVID-19 contact tracing efforts are within the realm of current CSP's privacy policies and if there have been any recent updates to privacy policies that would make COVID-19 contact tracing possible. We will also address necessary adjustments to policy if any, that will help achieve legislation compliant COVID-19 contact tracing applications. 

\section{Methods}
\label{sec:method}
To answer the research questions proposed in this study (mentioned in section~\ref{sec:intro}) and to compare and contrast each data privacy legislation and CSP privacy policies, we read through the publicly available laws and privacy policies of six privacy legislation (GDPR, LGPD, COPPA, HIPPA, CCPA, and SHIELD) and five CSPs' privacy policies (Verizon Wireless, AT\&T Mobility, T-Mobile US, Boost Mobile, and U.S. Cellular).

During each initial read through, all pertinent information regarding the usage, processing, and protection of mobile location data (MLD), was noted. With regard to the legislation, this information included MLD definitions under the law and the protections afforded to MLD. As for CSP privacy policies, this included information on how MLD is collected, processed, protected, and distributed. After the initial data collection, a second read through was performed to insure no vital information was missed. Finally, a thematic analysis was conducted. An inductive approach was adopted, by developing common themes found amongst every data privacy regulation and CSP privacy policy. 

To get a better understanding of the regulations that are being enforced in the United States and across the world in general, six privacy regulatory laws were analyzed, two of which are at the state level in the United States and four at the Global level.

\subsubsection{State level}
\begin{itemize}
\item California: CCPA~\cite{CCPA_2018}
\item New York: SHIELD Act~\cite{SHIELD_2019}
\end{itemize}

\subsubsection{Global level}
\begin{itemize}
\item EU: GDPR~\cite{GDPR_2019}
\item Brazil: LGPD~\cite{LGPD_2019}
\item United States: COPPA~\cite{COPPA_2020}
\item United States: HIPAA~\cite{HIPAA_2013}
\end{itemize}

We then analyzed privacy policies of the CSPs. Five of the largest CSPs in the United States were selected, based on the number of data subjects each of them has: Verizon Wireless~\cite{verizon_2020},  AT\&T Mobility~\cite{att_2020}, T-Mobile US~\cite{t-mobile_2020}, Boost Mobile~\cite{boost_mobile_2019} and U.S. Cellular~\cite{u.s._cellular_2020}. 
We collected the official policies of the six legislations as mentioned above. Thereafter, we focused on individual sections, which detail mobile data and/or user location data. For example GDPR while discussing Location access data management mentions that: ~\lq\lq The data subject should have the right not to be subject to a decision... in particular to analyse or predict aspects concerning the data subject’s ... location or movements...\rq\rq~.\\

Similarly, LGPD which takes a lot of content from the GDPR, notes that~\lq\lq Activities of processing of personal data shall be done in good faith \rq\rq~ this includes location data as it is implied in the personal data definition. Additionally, other privacy policies such as, CCPA details the mobile data location access strategies as, ~\lq\lq collection, use, retention, and sharing of a consumer’s personal information shall be reasonably necessary and proportionate...\rq\rq~ location data is explicitly defined as personal information in THE CCPA regulation. 

Amongst the other privacy policies, COPPA primarily focuses on Children Location Data, and thus while addressing several concerns mentions that operators must, ~\lq\lq  establish and maintain reasonable procedures to protect the confidentiality, security, and integrity of personal information collected from children \rq\rq~. HIPAA primarily focuses on an individual's health related data. Thus, for our study it was interesting to see how contact-tracing applications, built to stop the COVID-19 spread address the HIPAA concerns. It mentions~\lq\lq The provider may then disclose the individual’s condition and location in the facility to anyone asking for the individual by name \rq\rq~.

\begin{table*}[h]
  \centering
  \begin{tabular}{|l|p{1cm}|p{1cm}|p{1cm}|p{1cm}|p{1.5cm}|}
   \cline{2-6}
     \multicolumn{1}{c|}{} & Geo-location defined as PI & Geo-location implied as PI & Right to know & Right to delete & Right to opt-out\\ 
    \hline
    Data Privacy Regulations & Three Legislation (COPPA, GDPR, CCPA) & Two Legislation (SHIELD, LGPD) & Four Legislation (LGPD, COPPA, GDPR, CCPA) & Four Legislation (LGPD, COPPA, GDPR, CCPA) & Four Legislation (LGPD, COPPA, GDPR, CCPA)\\ \hline
    CSP Privacy Policy & All Five CSPs & All Five CSPs & All Five CSPs & One CSP (U.S. Cellular) & Four CSPs (Verizon Wireless,  T-mobile US, Boost Mobile, U.S. Cellular)  \\ \hline
  \end{tabular}
    \vspace{5mm}
  \caption{Privacy Protection Features Mentioned Across Data Privacy Legislation (Row 1) and Compliance of these Privacy-focused Featured Within Privacy Policies Defined by the Cellular Service Providers (Row2).}
  \label{tab:Themes}
\end{table*}
\begin{table*}[h]
  \centering
  \begin{tabular}{|p{0.5cm}|p{1.45cm}|p{1.3cm}|p{1.3cm}|p{1.3cm}|p{1.3cm}|p{1.3cm}|}
  \cline{2-7}
  \multicolumn{1}{c|}{} &
    \multicolumn{4}{c|}{Automatically} & \multicolumn{2}{c|}{Provided} \\ 
    \cline{2-7}
  \multicolumn{1}{c|}{} &
    \multicolumn{4}{c|}{4} & \multicolumn{2}{c|}{3} \\ 
    \cline{2-7}
  \multicolumn{1}{c|}{} &Cellphone Towers & Wi-Fi & Bluetooth & GPS & Zip code & Consumer Home Address\\ 
    \hline
    & & & & & & \\ 
     Number of CSPs & Three CSPs (AT\&T Mobility, Verizon Wireless, U.S. Cellular)  & Two CSPs (AT\&T Mobility, Verizon Wireless) & Two CSPs (AT\&T Mobility, T-Mobile) & Two CSPs (AT\&T Mobility, Verizon Wireless) & Three CSPs (AT\&T Mobility, US Cellular, & Three CSPs (AT\&T Mobility, US Cellular)\\ \hline
  \end{tabular}
    \vspace{5mm}
  \caption{Types of Data Collection Methods Applied By the Cellphone Service Providers to Collect Consumer Mobile Location Data}
  \label{tab:How is MLD being collected by CSPs}
\end{table*}

\subsection{Analysis}
Thematic Analysis
After reading through each regulation individually, five common themes were found: 
\begin{itemize}
    \item geo-location is explicitly defined as personal information (PI)
    \item geo-location is implied as PI
    \item the data subjects are granted the right to know
    \item the data subjects are granted the right to delete
    \item the data subjects are granted the right to opt-out
\end{itemize}

HIPAA did not provide any substantial policies that involve MLD, rather HIPAA only protects the location of a patient inside a hospital such as a room number or ward~\cite{HIPAA_2013}. HIPPA does not provide substantial information for the purposes of this study, Therefore, it is left out of our final analysis and results.

We conducted a thematic analysis on the privacy policies of the five CSPs selected. We concentrated on the methods they used to gather data and found common themes between all the CSPs:
\begin{enumerate}
    \item Automatic data collection: Cell Towers, WiFi, Bluetooth and GPS.

    \item User provided: Zip Code and Home address.

\end{enumerate}
Additionally, we found common themes between these CSPs describing the entities that have access to mobile location data collected: Third parties, Emergency or legal services, and Account holder.
    
However, we were unable to analyze the methods employed by the CSPs to protect the mobile location data as the policies provided vague information on these methods. Some of the details in protecting user data included prevention and mitigation concepts such as: \textit{Authentication} and \textit{Incident response plans or} safeguards.

\section {Results}
\label{sec:results}
After discarding HIPAA from the thematic analysis, we were left with five data protection legislations: GDPR, LGPD, CCPA, COPPA, and SHIELD. While not all definitions of PI are uniform across the legislation, MLD can be interpreted as PI under each legislation. 
Three of the five legislations explicitly included MLD in private information definition, whereas only two legislations include verbiage which defines PI as "information regarding any identified or 'identifiable' natural person"~\cite{SHIELD_2019}. Despite the lack of an explicit definition in New York's SHIELD act and the LGPD, MLD is protected under every legislation analyzed in this study as a private information (PI). 

Of the 5 legislations analyzed, only one (SHIELD act) does not protect the right for a data subject to know what PI is being collected, four legislations (LGPD, COPPA, GDPR, CCPA) protect the subjects' rights to delete PI, as for the right to opt-out, it is protected under four (LGPD, COPPA, GDPR, CCPA) legislations. These results can be seen in table~\ref{tab:Themes}.

In our analysis of the CSP privacy policies, we found that all five CSPs defined MLD as PI, all five complied with the data subjects' rights to know, and only one CSP (AT\&T Mobility) did not comply with the data subjects' rights to opt-out. However, compliance with a data subjects' rights to delete was less straightforward. In fact, only one CSP provided the option to delete PI to all data subjects, where as the remaining four CSPs only provided the right to delete to California residents as this is required under the CCPA~\cite{CCPA_2018}. Two (AT\&T Mobility, Boost Mobile) of these four CSPs also provides the opportunity to delete data of any data subject under the age of 13, thus complying with COPPA regulation~\cite{COPPA_2020}. As such only three (AT\&T Mobility, Boost Mobile, U.S. Cellular) of the five CSPs complied with COPPA's right to delete.

Variation among CSP privacy policies are most visible in their compliance with the legislations, as well as the methods they collect mobile location data. When addressing how MLD is collected, all five CSPs collect MLD automatically, but the privacy policies phrase the collection methods differently. Some CSPs use different means to collect MLD. As shown in Table~\ref{tab:How is MLD being collected by CSPs}, three CSPs (AT\&T Mobility, Verizon Wireless, U.S. Cellular) collect MLD through cell towers, two (AT\&T Mobility, Verizon Wireless) collect it through Wi-Fi access points, two  (AT\&T Mobility, T-Mobile) collect through Bluetooth, and two collect it using GPS (AT\&T Mobility, Verizon Wireless).
While these results are not representative of all the automatic MLD collection methods stated by the CSPs' privacy policies, it is important to understand that this analysis was done solely on what is written in each privacy policy. In fact, Two of the five privacy policies analyzed did not elaborate on how MLD is automatically collected (Boost Mobile, T-Mobile) as such we were unable to determine if they used any of the four most common methods. Additionally, two CSPs policies, Verizon Wireless and AT\&T Mobility, state that they use all four methods of automatic collection~\cite{verizon_2020}~\cite{att_2020}. Three privacy policies stated that MLD is provided by the account holder, and each policy stated that this occurs when creating an account for a data subject. 

\begin{table*}[t]
  \centering
  \begin{tabular}{|p{1cm}|p{1.5cm}|p{1.5cm}|p{1.5cm}|p{2.5cm}|p{1.5cm}|}
  \cline{2-6}
  \multicolumn{1}{c|}{} &
    Authentication & Incident Response Plan/Safeguards & Retained only as long as needed & Anonymous aggregate reports & Employee Training\\
 \hline
     Number of CSPs & Four CSPs (AT\&T Mobility, T-Mobile, Boost Mobile, U.S. Cellular) & Two CSPs (Boost Mobile, U.S. Cellular) & Two CSPs (Verizon Wireless, U.S. Cellular) & Four CSPs (Verizon Wireless, T-Mobile, Boost Mobile, U.S. Cellular)  & Two CSPs (Verizon Wireless, U.S. Cellular)\\ \hline
  \end{tabular}
  \vspace{5mm}
  \caption{Security Measures Implemented by the Cellphone Service Providers to Process and Protect Consumer Mobile Location Data as Mentioned in the Privacy Policies}
  \label{tab:MethodsCSP}
\end{table*}

\begin{table*}[t]
  \centering
  \begin{tabular}{|p{1.75cm}|p{1.75cm}|p{4.5cm}|p{3cm}|}
  \cline{2-4}
  \multicolumn{1}{c|}{} &
    Third Parties & Emergency/Legal Services & Account Holder \\ 
    \hline
    Number of CSPs & All Five CSPs & Four CSPs (Verizon Wireless, T-Mobile, Boost Mobile, U.S. Cellular) & Two CSPs (T-Mobile, U.S. Cellular)\\ \hline
  \end{tabular}
    \vspace{5mm}
      \caption{Data Access Provided to the Mobile Located Data as collected by the CSPs Starting from Third Parties, Emergency/Legal Services, and Account Holder as Mentioned in the Privacy Policies.}
  \label{tab:accessentity}
\end{table*}
Privacy policies stating how MLD is protected was found less frequently, with two CSPs (Verizon Wireless, U.S. Cellular) retaining MLD only as long as needed and four CSPs (Verizon Wireless, T-Mobile, Boost Mobile, U.S. Cellular) creating anonymous aggregate reports of the MLD. The ways in which privacy policies explain how CSPs protect MLD was less thorough with four (AT\&T Mobility, T-Mobile, Boost Mobile, U.S. Cellular) requiring some form of authentication to access MLD, two (Boost Mobile, U.S. Cellular) creating "incident response plans or safeguards", and two CSPs use employee training to protect MLD (Verizon Wireless, U.S. Cellular). These results can be seen in Table~\ref{tab:MethodsCSP}.

When it comes to providing access to MLD, all five CSPs provide third parties with data subject's MLD. Only one CSP (AT\&T Mobility) did not mention in their privacy policy that they provide MLD to emergency or legal services upon request. Two policies (T-Mobile, U.S. Cellular) state that the account holder is also given access to all MLD collected by the CSP. These results are shown in Table~\ref{tab:accessentity}.

\section{Discussion}
\label{sec:discussion}
Through the thematic analysis of data privacy regulations and CSP privacy policies, we were able to answer the five proposed research questions of this study. First, whether CSP privacy policies are compliant with current data protection regulations, cannot be simply answered with a yes or a no but rather addressed as individual clauses of each regulation.

All five CSPs explicitly define MLD as Personal Information (PI) which is in agreement with all five data protection regulation's definitions. As required by four data privacy regulations (GDPR, LGPD, COPPA and CCPA), the right to know what data is collected is properly addressed by all five CSPs. Similarly, these four data privacy regulations require, the right to opt-out is adhered to by four CSPs (Verizon Wireless, T-Mobile, Boost Mobile and U.S. Cellular). In this sense, CSP privacy policies are majorly in compliance with current data privacy regulations. However, the right to delete MLD is only observed by Verizon, while four CSPs (At\&T Mobility, T-Mobile USm Boost Mobile and U.S. Cellular) only afforded that right to Californian residents and as such are in violation of GDPR~\cite{GDPR_2019} and LGPD~\cite{LGPD_2019} legislation.

In late 2019, the United States Congress introduced the Consumer Online Rights Privacy Act (COPRA) bill. Under the current version of this bill, MLD is defined as ~\lq\lq sensitive covered data\rq\rq~and also protects the right to know, to delete, and to opt-out~\cite{CORPA_2019}. Similar to other national data protections in the US, the Federal Trade Commission (FTC) would then develop a department to enforce this proposed act. The development of a national legislation in the United States will likely rectify this disparity between rights for California residents and the rest of US citizens. 

To answer the second question of how MLD is collected, we found that automated processes are in use most frequently, and often are found in a combination of cell tower data, Wi-Fi access point data, Bluetooth, and GPS data, where cell tower data is the most common form of data collection. Less frequently, MLD is provided by a data subject, however it is important to note that this location data is usually in the form of a data subject's zip code and home address and is not active MLD (precise location). It is vital that this information is included in a CSP's privacy policy, because the majority of MLD is collected in an automated way, and it is often easy for consumers to forget about these processes. Furthermore, MLD received by most CSPs is anonymized in aggregate reports, but fewer CSPs retained the MLD only as long as needed , which can be concerning. In fact, privacy policies often do not go into specifics in regards to data processing or protection, as such it is difficult for the users to assess how effective and thorough any of the details mentioned in the privacy policies are. 

Transparency in privacy policies is extremely important and is even mandated under recital 58 of the GDPR, such that policies must be ~\lq\lq concise, easily accessible and easy to understand, and that clear and plain language [...] be used\rq\rq~~\cite{GDPR_2019}. All the CSPs' privacy policies that were evaluated adhered to this requirement, however in an area such as MLD, where collection and processes may not be easily understandable to the average consumer, vital information about collection and storage processes should be explicitly mentioned, which none of the CSPs did. In this case, only one privacy policy (Verizon Wireless) included a notice dedicated to describing the collection and processing of MLD~\cite{verizon_mobile_location_analytics_2019}, as such it is clear that this level of transparency is not a common practice for American CSPs. 

Finally, within the scope of our present user monitoring to prevent the spread of COVID-19, none of the CSP privacy policies explicitly detail the purpose of the MLDs collected or mention the change of location data collection due to the situation. However, as each policy stands, it is possible that contact tracing could be developed without any policy change. All but one CSP privacy policy analyzed in this study state that they will provide MLD to emergency or legal services if requested. Currently, this is being used primarily to convict criminals by placing them at the scene of a crime, or by emergency response teams to locate a person in need of help ~\footnote{https://cacm.acm.org/magazines/2020/10/247585-who-has-access-to-your-smartphone-data/fulltext}. However, COVID-19 has been considered as a national sanitary emergency by many governments~\footnote{https://www.who.int/publications/m/item/COVID-19-public-health-emergency-of-international-concern-(pheic)-global-research-and-innovation-forum}, therefore the collection of MLD by emergency services for contact tracing purposes is within the scope of these privacy policies. 

CSPs also state in their privacy policies that MLD is shared with partnered third parties, this statement itself is vague and could also allow for CSPs to partner with future apps or companies whose purpose is to use MLD for variety of purposes. Another way that CSPs could access MLD for contact tracing purposes, is simply through data subject consent. If CSPs wish to use MLD for contact tracing, it would be most advantageous and regulation compliant to adjust their privacy policy in order to keep each data subject informed. Thus, we see that there are several consistency issues when it comes to CSPs privacy policies with the data collection. Thus, we provide a few recommendations obtained through our research in the following section.
\section{Implications}
\label{sec:implications}
In this section we will address the results and observations from the analysis of our study in the realm of data privacy frameworks, analyses of location privacy laws, and COVID-19 contact tracing. We will also provide actionable recommendations for the users, developers, policy makers, and organizations, as well as policy change or update suggestions, based on our findings from this study and several other studies.
\subsection{Data privacy frameworks}
In our study, we did not develop a new data privacy framework, however we took into consideration many different privacy frameworks such as the ones introduced by Ahamed et al.~\cite{AhamedSheikh2012Anlp}, Lee et al.~\cite{LeeChao2013AFoE}, and primarily Cavoukian~\cite{cavoukian_2011} work and conducted a comparative analysis.

Understanding how policies are written can help develop frameworks that focus on the policies rather than on procedures. Implementing an effective and efficient privacy policy is key~\cite{das2019privacy}. In this study, we found that there was a lack of transparency when it came to actual procedures followed by the CSPs. As such a modular approach to  transparency in privacy policy would provide a huge improvement to current policies~\cite{das2018modularity}. We also recommend introducing the policy in plain, easy to understand language as Kumaraguru et al. suggests~\cite{kumaraguru2007survey}, 
using easily understandable language is critical~\cite{sadeh2013usable}. This level of transparency would make it easier for common users to have a better understanding of the implications of using a specific CSP~\cite{jensen2004privacy}, as well as make it easier for scholars to conduct research and to determine if actual practices align with policy~\cite{das2020humans}.

\subsection{Analysis of location privacy laws}
In our analysis of six data privacy legislation, we found that most regulations have similar requirements for the protection of MLD. Understanding that CSP privacy policies are mostly compliant with the current leading legislation is important because it provides more opportunities to develop widespread and helpful legislation in the US. Implementing new national policy in the US will continue to enforce the current model of CSP privacy policies and lead to the development of better consumer data protections~\cite{harding2019understanding,reichert2020privacy}. 

\subsection{COVID-19 contact tracing efforts of MLD}
The works of Egan and Oliver et al. discuss the importance of COVID-19 contact tracing and the privacy concerns that are brought to the table if MLD were used for contact tracing efforts. We were able to contribute to their works by analyzing CPS privacy policies and data privacy regulations to determine if privacy concerns are viable in the instance related to COVID-19 contact tracing.

Due to the lack of CSPs' privacy policy transparency~\cite{bengio2021inherent}, we were  unable to collect enough data, which in turn made it impossible to gain an in-depth understanding of the CSPs' mobile location data collection and use processes. We are however able to assert that it is possible for CSPs to use MLD for COVID-19 contact tracing within their privacy policies' guidelines. Nonetheless we recommend that the CSPs update their privacy policy if they intend to use MLD for contact tracing. Proper security measures should also be adopted when using contact tracing, such as encryption, anonymization, and obfuscation~\cite{hawkes2007method,baumgartner2020mind,gvili2020security}.  At the time of this research, we are not aware of any privacy policy changes made by CSPs to allow for COVID-19 contact tracing. An update describing the extant of MLD use in contact tracing efforts and preferably the opportunity for a data subject to opt-out, would reflect positively on the CSPs~\cite{springer2020progressive,kleinman2020digital}.

\section {Future Work and Limitations}
Using the methodology explained in this study, we believe it is important to analyze more CSPs. This should provide a better understanding of CSP privacy policy on the macro scale. 
It will also be advantageous for future extension of this work to dig deeper into the actual processes which CSPs use to collect, process, protect, and distribute MLD rather than just analyzing what is mentioned by them in the privacy policies. A thorough understanding of what processes are used to collect MLD, how and if MLD is actually anonymized, and what reports third parties and emergency or legal services receive through the CSPs, will help to build the bigger picture of how CSPs interact and utilize with the MLD they collect. Finally, we believe it is critical to develop technologies that comply with data privacy regulations. Any future works on COVID-19 contact tracing efforts will be able to utilize the findings of the study to determine how best to comply with legal regulations when interacting with MLD.   

\section {Conclusion}
\label{sec:conclusion}
Since the introduction of the GDPR in 2016, the rise of data protection regulations have become a notable force that should be addressed by privacy policies of all organisations that handle Personally Identifiable Information (PII). In addition, the development of the COVID-19 pandemic demands unique solutions to prevent further spread of the virus, such as contact tracing applications which monitors the location of users. Such changes, demand to see whether the CSPs are adhering to the privacy policies and user perceptions when it comes to handling critical data such as precise location. 

In this regard, the contribution of this study is to get a primary understanding of five big CSP's privacy policies as they relate to the collection, processing, protection, and sharing of mobile location data under the scope of data protection regulations. This study analyzes state-wide, national, and international data privacy regulations (SHIELD ACT, CCPA, HIPAA, COPPA, LGPD, and GDPR) with the privacy policies of five most used cellphone provider services (Verizon Wireless, AT\&T Mobility, T-Mobile US, Boost Mobile, and U.S. Cellular) in the United States is very timely and critical. 

Our study provides us with a lens to better determine the viability of COVID-19 contact tracing under current CSP privacy policies. While some contact tracing may be possible, we determined it is likely necessary that changes to policy and appropriate measures to protect MLD should be implemented. This study was limited in analyzing exclusively what was stated in each privacy policy rather than the true practices of CSPs which is mentioned as the future extension of this work. Further contributions should expand by analyzing additional CSPs privacy policies, analyzing actual processes used by CSPs to process and share MLD, look into further policy changes which allow the use of MLD for COVID-19 contact tracing while minimizing privacy violations, or developing regulation compliant technologies for contact tracing. We conclude by providing actionable recommendations for users, policy makers, developers, and organizations.

\section*{Acknowledgments}
We would like to acknowledge the Security and Privacy Research Lab at the University of Denver and the students of the COMP 3705/4705 : Adv Topics: Human-Centered Data Security and Privacy for their initial feedback. Any opinions, findings, and conclusions or recommendations expressed in this material are solely those of the authors and do and do not necessarily reflect the views of the University of Denver.
\bibliographystyle{IEEEtran}
\bibliography{references}

\begin{thebibliography}{10}
\providecommand{\url}[1]{#1}
\csname url@samestyle\endcsname
\providecommand{\newblock}{\relax}
\providecommand{\bibinfo}[2]{#2}
\providecommand{\BIBentrySTDinterwordspacing}{\spaceskip=0pt\relax}
\providecommand{\BIBentryALTinterwordstretchfactor}{4}
\providecommand{\BIBentryALTinterwordspacing}{\spaceskip=\fontdimen2\font plus
\BIBentryALTinterwordstretchfactor\fontdimen3\font minus
  \fontdimen4\font\relax}
\providecommand{\BIBforeignlanguage}[2]{{%
\expandafter\ifx\csname l@#1\endcsname\relax
\typeout{** WARNING: IEEEtran.bst: No hyphenation pattern has been}%
\typeout{** loaded for the language `#1'. Using the pattern for}%
\typeout{** the default language instead.}%
\else
\language=\csname l@#1\endcsname
\fi
#2}}
\providecommand{\BIBdecl}{\relax}
\BIBdecl

\bibitem{martin2020data}
K.~D. Martin, J.~J. Kim, R.~W. Palmatier, L.~Steinhoff, D.~W. Stewart, B.~A.
  Walker, Y.~Wang, and S.~K. Weaven, ``Data privacy in retail,'' \emph{Journal
  of Retailing}, vol.~96, no.~4, pp. 474--489, 2020.

\bibitem{10.1145/2981547}
\BIBentryALTinterwordspacing
N.~Saxena, S.~Grijalva, and N.~S. Chaudhari, ``Authentication protocol for an
  iot-enabled lte network,'' vol.~16, no.~4, 2016. [Online]. Available:
  \url{https://doi.org/10.1145/2981547}
\BIBentrySTDinterwordspacing

\bibitem{streiff2019overpowered}
J.~Streiff, S.~Das, and J.~Cannon, ``Overpowered and underprotected toys
  empowering parents with tools to protect their children,'' in \emph{2019 IEEE
  5th International Conference on Collaboration and Internet Computing
  (CIC)}.\hskip 1em plus 0.5em minus 0.4em\relax IEEE, 2019, pp. 322--329.

\bibitem{hadan2019making}
H.~Hadan, N.~Serrano, S.~Das, and L.~J. Camp, ``Making iot worthy of human
  trust,'' \emph{Available at SSRN 3426871}, 2019.

\bibitem{gopavaram2019iotmarketplace}
S.~R. Gopavaram, J.~Dev, S.~Das, and J.~Camp, ``Iotmarketplace: Informing
  purchase decisions with risk communication,'' 2019.

\bibitem{yang2010determinants}
K.~Yang, ``Determinants of us consumer mobile shopping services adoption:
  implications for designing mobile shopping services,'' \emph{Journal of
  consumer marketing}, 2010.

\bibitem{hatt_jarich_2019}
T.~Hatt and P.~Jarich, ``Global mobile trends 2020,'' \url
  {https://data.gsmaintelligence.com/research/research/research-2019/global-mobile-trends-2020},
  2019.

\bibitem{petzer2011perceived}
D.~Petzer and C.~De~Meyer, ``The perceived service quality, satisfaction and
  behavioural intent towards cellphone network service providers: A
  generational perspective,'' \emph{African journal of business management},
  vol.~5, no.~17, pp. 7461--7473, 2011.

\bibitem{tsai2010location}
J.~Y. Tsai, P.~G. Kelley, L.~F. Cranor, and N.~Sadeh, ``Location-sharing
  technologies: Privacy risks and controls,'' \emph{Isjlp}, vol.~6, p. 119,
  2010.

\bibitem{sung2020zipphone}
K.~Sung, B.~Levine, and M.~Zheleva, ``Zipphone: Protecting user location
  privacy from cellular service providers,'' \emph{arXiv preprint
  arXiv:2002.04731}, 2020.

\bibitem{case319cv04063}
\emph{Case No. 19-cv-4063, Scott, Jewel, And Pontis, et al. v. AT\&T Inc.;
  AT\&T Services, Inc.; AT\&T Mobility, LLC; Technocom Corp.;and Zumigo, Inc.}

\bibitem{khan2020spread}
S.~Khan, R.~Siddique, A.~Ali, Q.~Bai, Z.~Li, H.~Li, M.~A. Shereen, M.~Xue, and
  G.~Nabi, ``The spread of novel coronavirus has created an alarming situation
  worldwide,'' \emph{Journal of infection and public health}, vol.~13, no.~4,
  p. 469, 2020.

\bibitem{das2020change}
S.~Das, A.~Kim, and S.~Karmakar, ``Change-point analysis of
  cyberbullying-related twitter discussions during covid-19,'' \emph{arXiv
  preprint arXiv:2008.13613}, 2020.

\bibitem{karmakar2020evaluating}
S.~Karmakar and S.~Das, ``Evaluating the impact of covid-19 on cyberbullying
  through bayesian trend analysis,'' in \emph{Proceedings of the European
  Interdisciplinary Cybersecurity Conference}, 2020, pp. 1--6.

\bibitem{karmakar2021understanding}
------, ``Understanding the rise of twitter-based cyberbullying due to covid-19
  through comprehensive statistical evaluation,'' in \emph{Proceedings of the
  54th Hawaii International Conference on System Sciences}, 2021.

\bibitem{jenkins2021portrait}
E.~K. Jenkins, C.~McAuliffe, S.~Hirani, C.~Richardson, K.~C. Thomson,
  L.~McGuinness, J.~Morris, A.~Kousoulis, and A.~Gadermann, ``A portrait of the
  early and differential mental health impacts of the covid-19 pandemic in
  canada: findings from the first wave of a nationally representative
  cross-sectional survey,'' \emph{Preventive Medicine}, p. 106333, 2021.

\bibitem{eames2003contact}
K.~T. Eames and M.~J. Keeling, ``Contact tracing and disease control,''
  \emph{Proceedings of the Royal Society of London. Series B: Biological
  Sciences}, vol. 270, no. 1533, pp. 2565--2571, 2003.

\bibitem{keeling2020efficacy}
M.~J. Keeling, T.~D. Hollingsworth, and J.~M. Read, ``Efficacy of contact
  tracing for the containment of the 2019 novel coronavirus (covid-19),''
  \emph{J Epidemiol Community Health}, vol.~74, no.~10, pp. 861--866, 2020.

\bibitem{lalmuanawma2020applications}
S.~Lalmuanawma, J.~Hussain, and L.~Chhakchhuak, ``Applications of machine
  learning and artificial intelligence for covid-19 (sars-cov-2) pandemic: A
  review,'' \emph{Chaos, Solitons \& Fractals}, p. 110059, 2020.

\bibitem{rorres2018contact}
C.~Rorres, M.~Romano, J.~A. Miller, J.~M. Mossey, T.~H. Grubesic, D.~E.
  Zellner, and G.~Smith, ``Contact tracing for the control of infectious
  disease epidemics: Chronic wasting disease in deer farms,'' \emph{Epidemics},
  vol.~23, pp. 71--75, 2018.

\bibitem{hebert2020beyond}
L.~H{\'e}bert-Dufresne, B.~M. Althouse, S.~V. Scarpino, and A.~Allard, ``Beyond
  r0: The importance of contact tracing when predicting epidemics,''
  \emph{medRxiv}, 2020.

\bibitem{sahu2020covid}
K.~K. Sahu, A.~K. Mishra, and A.~Lal, ``Covid-2019: update on epidemiology,
  disease spread and management,'' \emph{Monaldi Archives for Chest Disease},
  vol.~90, no.~1, 2020.

\bibitem{baumgartner2020mind}
L.~Baumg{\"a}rtner, A.~Dmitrienko, B.~Freisleben, A.~Gruler, J.~H{\"o}chst,
  J.~K{\"u}hlberg, M.~Mezini, M.~Miettinen, A.~Muhamedagic, T.~D. Nguyen
  \emph{et~al.}, ``Mind the gap: Security \& privacy risks of contact tracing
  apps,'' \emph{arXiv preprint arXiv:2006.05914}, 2020.

\bibitem{parker2020ethics}
M.~J. Parker, C.~Fraser, L.~Abeler-D{\"o}rner, and D.~Bonsall, ``Ethics of
  instantaneous contact tracing using mobile phone apps in the control of the
  covid-19 pandemic,'' \emph{Journal of Medical Ethics}, vol.~46, no.~7, pp.
  427--431, 2020.

\bibitem{cho2020contact}
H.~Cho, D.~Ippolito, and Y.~W. Yu, ``Contact tracing mobile apps for covid-19:
  Privacy considerations and related trade-offs,'' \emph{arXiv preprint
  arXiv:2003.11511}, 2020.

\bibitem{zeinalipour2020covid}
D.~Zeinalipour-Yazti and C.~Claramunt, ``Covid-19 mobile contact tracing apps
  (mcta): A digital vaccine or a privacy demolition?'' in \emph{2020 21st IEEE
  International Conference on Mobile Data Management (MDM)}.\hskip 1em plus
  0.5em minus 0.4em\relax IEEE, 2020, pp. 1--4.

\bibitem{shukla2020privacy}
M.~Shukla, S.~Lodha, G.~Shroff, R.~Raskar \emph{et~al.}, ``Privacy guidelines
  for contact tracing applications,'' \emph{arXiv preprint arXiv:2004.13328},
  2020.

\bibitem{bengio2021inherent}
Y.~Bengio, D.~Ippolito, R.~Janda, M.~Jarvie, B.~Prud'homme, J.-F. Rousseau,
  A.~Sharma, and Y.~W. Yu, ``Inherent privacy limitations of decentralized
  contact tracing apps,'' \emph{Journal of the American Medical Informatics
  Association}, vol.~28, no.~1, pp. 193--195, 2021.

\bibitem{bygrave2010privacy}
L.~A. Bygrave, ``Privacy and data protection in an international perspective,''
  \emph{Scandinavian studies in law}, vol.~56, no.~8, pp. 165--200, 2010.

\bibitem{mangini2020empirical}
V.~Mangini, I.~Tal, and A.-N. Moldovan, ``An empirical study on the impact of
  gdpr and right to be forgotten-organisations and users perspective,'' in
  \emph{Proceedings of the 15th International Conference on Availability,
  Reliability and Security}, 2020, pp. 1--9.

\bibitem{gdpr2018art4}
``Art 4. \uppercase{GDPR} - definitions,''
  \url{https://gdpr-info.eu/art-4-gdpr/}, Mar 2018.

\bibitem{LGPD_2019}
``Brazilian general data protection law (\uppercase{LGPD}, english
  translation),''
  \url{https://iapp.org/resources/article/brazilian-data-protection-law-lgpd-english-translation/},
  2019.

\bibitem{erickson2018comparative}
A.~Erickson, ``Comparative analysis of the eu's gdpr and brazil's lgpd:
  Enforcement challenges with the lgpd,'' \emph{Brook. J. Int'l L.}, vol.~44,
  p. 859, 2018.

\bibitem{gdpr2016art5}
``Art. 5 \uppercase{GDPR} – principles relating to processing of personal
  data,'' \url{https://gdpr-info.eu/art-5-gdpr/}, Aug 2016.

\bibitem{gdpr2018art25}
``Art. 25 \uppercase{GDPR} – data protection by design and by default,''
  \url{https://gdpr-info.eu/art-25-gdpr/}, Mar 2018.

\bibitem{gdpr2016art13}
``Art. 13 \uppercase{GDPR} – information to be provided where personal data
  are collected from the data subject,''
  \url{https://gdpr-info.eu/art-13-gdpr/}, Aug 2016.

\bibitem{9305893}
M.~{Bano}, D.~{Zowghi}, and C.~{Arora}, ``Requirements, politics, or
  individualism: What drives the success of covid-19 contact-tracing apps?''
  \emph{IEEE Software}, vol.~38, no.~1, pp. 7--12, 2021.

\bibitem{walrave2020ready}
M.~Walrave, C.~Waeterloos, and K.~Ponnet, ``Ready or not for contact tracing?
  investigating the adoption intention of covid-19 contact-tracing technology
  using an extended unified theory of acceptance and use of technology model,''
  \emph{Cyberpsychology, Behavior, and Social Networking}, 2020.

\bibitem{marcut2018analysis}
M.~Marcut, ``Analysis of gdpr implementation at county level,'' in
  \emph{Sustainable Development and Resilience of Local Communities and Public
  Sector Organizations. Conference Proceedings' Transylvanian International
  Conference in Public Administration}, 2018, pp. 16--18.

\bibitem{raposo2019lgpd}
C.~F.~L. Rap{\^o}so, H.~M. de~Lima, W.~F. de~Oliveira~Junior, P.~A.~F. Silva,
  and E.~E. de~Souza~Barros, ``Lgpd-lei geral de prote{\c{c}}{\~a}o de dados
  pessoais em tecnologia da informa{\c{c}}{\~a}o: Revis{\~a}o
  sistem{\'a}tica,'' \emph{RACE-Revista de Administra{\c{c}}{\~a}o do Cesmac},
  vol.~4, pp. 58--67, 2019.

\bibitem{rothstein2019california}
M.~A. Rothstein and S.~A. Tovino, ``California takes the lead on data privacy
  law,'' \emph{Hastings Center Report}, vol.~49, no.~5, pp. 4--5, 2019.

\bibitem{lance2019erecting}
L.~Lance Plunkett~JD, ``Erecting a shield against the bad guys,'' \emph{New
  York State Dental Journal}, vol.~85, no.~6, pp. 4--7, 2019.

\bibitem{reyes2018won}
I.~Reyes, P.~Wijesekera, J.~Reardon, A.~E.~B. On, A.~Razaghpanah,
  N.~Vallina-Rodriguez, and S.~Egelman, ``“won’t somebody think of the
  children?” examining coppa compliance at scale,'' \emph{Proceedings on
  Privacy Enhancing Technologies}, vol. 2018, no.~3, pp. 63--83, 2018.

\bibitem{assistance2003summary}
H.~C. Assistance, ``Summary of the hipaa privacy rule,'' \emph{Office for Civil
  Rights}, 2003.

\bibitem{liu2007data}
L.~Liu, ``From data privacy to location privacy: models and algorithms,'' in
  \emph{Proceedings of the 33rd international conference on Very large data
  bases}.\hskip 1em plus 0.5em minus 0.4em\relax Citeseer, 2007, pp.
  1429--1430.

\bibitem{beresford2003location}
A.~R. Beresford and F.~Stajano, ``Location privacy in pervasive computing,''
  \emph{IEEE Pervasive computing}, vol.~2, no.~1, pp. 46--55, 2003.

\bibitem{hoepman2014privacy}
J.-H. Hoepman, ``Privacy design strategies,'' in \emph{IFIP International
  Information Security Conference}.\hskip 1em plus 0.5em minus 0.4em\relax
  Springer, 2014, pp. 446--459.

\bibitem{AhamedSheikh2012Anlp}
S.~Ahamed, M.~Haque, and C.~Hasan, ``\BIBforeignlanguage{eng}{A novel location
  privacy framework without trusted third party based on location anonymity
  prediction},'' \emph{\BIBforeignlanguage{eng}{ACM SIGAPP Applied Computing
  Review}}, vol.~12, no.~1, pp. 24--34, 2012.

\bibitem{LeeChao2013AFoE}
C.~Lee, Y.~Guo, and L.~Yin, ``\BIBforeignlanguage{eng}{A framework of
  evaluation location privacy in mobile network},''
  \emph{\BIBforeignlanguage{eng}{Procedia computer science}}, vol.~17, pp.
  879--887, 2013.

\bibitem{shaham2020privacy}
S.~Shaham, M.~Ding, B.~Liu, S.~Dang, Z.~Lin, and J.~Li, ``Privacy preserving
  location data publishing: A machine learning approach,'' \emph{IEEE
  Transactions on Knowledge and Data Engineering}, 2020.

\bibitem{cavoukian_2011}
\BIBentryALTinterwordspacing
A.~Cavoukian, ``Privacy by design the 7 foundational principles,'' Jan 2011.
  [Online]. Available: \url{www.privacybydesign.ca}
\BIBentrySTDinterwordspacing

\bibitem{GeorgiadouYola2019LPit}
Y.~Georgiadou, R.~de~By, and O.~Kounadi, ``\BIBforeignlanguage{eng}{Location
  privacy in the wake of the gdpr},'' \emph{\BIBforeignlanguage{eng}{ISPRS
  international journal of geo-information}}, vol.~8, no.~3, pp. 1--17, 2019.

\bibitem{apthorpe2019evaluating}
N.~Apthorpe, S.~Varghese, and N.~Feamster, ``Evaluating the contextual
  integrity of privacy regulation: Parents' iot toy privacy norms versus
  $\{$COPPA$\}$,'' in \emph{28th $\{$USENIX$\}$ Security Symposium
  ($\{$USENIX$\}$ Security 19)}, 2019, pp. 123--140.

\bibitem{streiff2018s}
J.~Streiff, O.~Kenny, S.~Das, A.~Leeth, and L.~J. Camp, ``Who's watching your
  child? exploring home security risks with smart toy bears,'' in \emph{2018
  IEEE/ACM Third International Conference on Internet-of-Things Design and
  Implementation (IoTDI)}.\hskip 1em plus 0.5em minus 0.4em\relax IEEE, 2018,
  pp. 285--286.

\bibitem{liccardi2014can}
I.~Liccardi, M.~Bulger, H.~Abelson, D.~J. Weitzner, and W.~Mackay, ``Can apps
  play by the coppa rules?'' in \emph{2014 Twelfth Annual International
  Conference on Privacy, Security and Trust}.\hskip 1em plus 0.5em minus
  0.4em\relax IEEE, 2014, pp. 1--9.

\bibitem{AtaeiMehrnaz2018CwPL}
M.~Ataei, A.~Degbelo, C.~Kray, and V.~Santos,
  ``\BIBforeignlanguage{eng}{Complying with privacy legislation: From legal
  text to implementation of privacy-aware location-based services},''
  \emph{\BIBforeignlanguage{eng}{ISPRS international journal of
  geo-information}}, vol.~7, no.~11, p. 442, 2018.

\bibitem{harari2020process}
G.~M. Harari, ``A process-oriented approach to respecting privacy in the
  context of mobile phone tracking,'' \emph{Current opinion in psychology},
  vol.~31, pp. 141--147, 2020.

\bibitem{thaichon2016analysis}
P.~Thaichon, K.~Sharma, K.~Raina, and S.~Kapoor, ``Analysis of consumers’
  intention values in the choice of a mobile service provider,'' \emph{Asian
  Journal of Business Research ISSN}, vol.~6, no.~1, p. 2016, 2016.

\bibitem{whaiduzzaman2014measuring}
M.~Whaiduzzaman and A.~Gani, ``Measuring security for cloud service provider: A
  third party approach,'' in \emph{2013 International Conference on Electrical
  Information and Communication Technology (EICT)}.\hskip 1em plus 0.5em minus
  0.4em\relax IEEE, 2014, pp. 1--6.

\bibitem{CaseyEgan2020Mldt}
C.~Egan, ``\BIBforeignlanguage{eng}{Mobile location data tracking tools raise
  privacy questions amid covid-19},'' \emph{\BIBforeignlanguage{eng}{SNL Kagan
  Media \& Communications Report}}, 2020.

\bibitem{unacast}
\BIBentryALTinterwordspacing
``Privacy.'' [Online]. Available: \url{https://www.unacast.com/privacy}
\BIBentrySTDinterwordspacing

\bibitem{OliverNuria2020Mpdf}
N.~Oliver, B.~Lepri, H.~Sterly, R.~Lambiotte, S.~Deletaille, M.~De~Nadai,
  E.~Letouzé, A.~A. Salah, R.~Benjamins, C.~Cattuto, V.~Colizza, N.~de~Cordes,
  S.~P. Fraiberger, T.~Koebe, S.~Lehmann, J.~Murillo, A.~Pentland, P.~N. Pham,
  F.~Pivetta, J.~Saramäki, S.~V. Scarpino, M.~Tizzoni, S.~Verhulst, and
  P.~Vinck, ``\BIBforeignlanguage{eng}{Mobile phone data for informing public
  health actions across the covid-19 pandemic life cycle},''
  \emph{\BIBforeignlanguage{eng}{Science advances}}, vol.~6, no.~23, pp.
  eabc0764--eabc0764, 2020.

\bibitem{CCPA_2018}
``California consumer privacy act of 2018,''
  \url{http://leginfo.legislature.ca.gov/faces/codes_displayText.xhtml?division=3.},
  2018.

\bibitem{SHIELD_2019}
``\uppercase{NY} state senate bill s5575b,''
  \url{https://www.nysenate.gov/legislation/bills/2019/s5575}, Jul 2019.

\bibitem{GDPR_2019}
``\uppercase{GDPR} offical legal text,'' \url{https://gdpr-info.eu/}, Sep 2019.

\bibitem{COPPA_2020}
``Children's online privacy protection act of 1998,''
  \url{http://www.ftc.gov/ogc/coppa1.htm}, 2020.

\bibitem{HIPAA_2013}
H.~O. o.~t. Secretary and O.~f. C.~R. (OCR), ``Summary of the \uppercase{HIPAA}
  privacy rule,''
  \url{https://www.hhs.gov/hipaa/for-professionals/privacy/laws-regulations/index.html},
  Jul 2013.

\bibitem{verizon_2020}
``Get to know our privacy policies.''
  \url{https://www.verizon.com/about/privacy/}, 2020.

\bibitem{att_2020}
``Full privacy policy - at\&t people: Planet: Possibilities,''
  \url{https://about.att.com/csr/home/privacy/full_privacy_policy.html}, Jun
  2020.

\bibitem{t-mobile_2020}
``"t-mobile privacy policy",''
  \url{https://www.t-mobile.com/privacy-center/our-practices/privacy-policy},
  Aug 2020.

\bibitem{boost_mobile_2019}
B.~Mobile, ``Boost mobile privacy policy,''
  \url{https://www.boostmobile.com/about/legal/privacy-policy.html?INTNAV=BotNav\%3ALegal\%3APrivacy},
  Dec 2019.

\bibitem{u.s._cellular_2020}
``Privacy statement,'' \url{https://www.uscellular.com/privacy}, Sep 2020.

\bibitem{CORPA_2019}
M.~Cantwell, ``Text - s.2968 - 116th congress (2019-2020): Consumer online
  privacy rights act,''
  \url{https://www.congress.gov/bill/116th-congress/senate-bill/2968/text}, Dec
  2019.

\bibitem{verizon_mobile_location_analytics_2019}
``Mobile location analytics,''
  \url{https://www.verizon.com/about/privacy/mobile-location-analytics-privacy-notice},
  Oct 2019.

\bibitem{das2019privacy}
S.~Das, J.~Dev, and L.~J. Camp, ``Privacy preserving policy framework:
  User-aware and user-driven,'' \emph{Available at SSRN 3445942}, 2019.

\bibitem{das2018modularity}
S.~Das, J.~Dev, and K.~Srinivasan, ``Modularity is the key a new approach to
  social media privacy policies,'' in \emph{Proceedings of the 7th Mexican
  Conference on Human-Computer Interaction}, 2018, pp. 1--4.

\bibitem{kumaraguru2007survey}
P.~Kumaraguru, L.~Cranor, J.~Lobo, and S.~Calo, ``A survey of privacy policy
  languages,'' in \emph{Workshop on Usable IT Security Management (USM 07):
  Proceedings of the 3rd Symposium on Usable Privacy and Security, ACM}, 2007.

\bibitem{sadeh2013usable}
N.~Sadeh, A.~Acquisti, T.~D. Breaux, L.~F. Cranor, A.~M. McDonald, J.~R.
  Reidenberg, N.~A. Smith, F.~Liu, N.~C. Russell, F.~Schaub \emph{et~al.},
  ``The usable privacy policy project,'' in \emph{Technical report, Technical
  Report, CMU-ISR-13-119}.\hskip 1em plus 0.5em minus 0.4em\relax Carnegie
  Mellon University, 2013.

\bibitem{jensen2004privacy}
C.~Jensen and C.~Potts, ``Privacy policies as decision-making tools: an
  evaluation of online privacy notices,'' in \emph{Proceedings of the SIGCHI
  conference on Human Factors in Computing Systems}, 2004, pp. 471--478.

\bibitem{das2020humans}
S.~Das, R.~S. Gutzwiller, R.~D. Roscoe, P.~Rajivan, Y.~Wang, L.~Jean~Camp, and
  R.~Hoyle, ``Humans and technology for inclusive privacy and security,'' in
  \emph{Proceedings of the Human Factors and Ergonomics Society Annual
  Meeting}, vol.~64, no.~1.\hskip 1em plus 0.5em minus 0.4em\relax SAGE
  Publications Sage CA: Los Angeles, CA, 2020, pp. 461--464.

\bibitem{harding2019understanding}
E.~L. Harding, J.~J. Vanto, R.~Clark, L.~Hannah~Ji, and S.~C. Ainsworth,
  ``Understanding the scope and impact of the california consumer privacy act
  of 2018,'' \emph{Journal of Data Protection \& Privacy}, vol.~2, no.~3, pp.
  234--253, 2019.

\bibitem{reichert2020privacy}
L.~Reichert, S.~Brack, and B.~Scheuermann, ``Privacy-preserving contact tracing
  of covid-19 patients.'' \emph{IACR Cryptol. ePrint Arch.}, vol. 2020, p. 375,
  2020.

\bibitem{hawkes2007method}
P.~M. Hawkes, R.~T. Hsu, R.~Rezaiifar, G.~G. Rose, P.~E. Bender, J.~Wang, R.~F.
  Quick~Jr, A.~C. Mahendran, and P.~A. Agashe, ``Method and apparatus for
  security in a data processing system,'' Feb.~27 2007, uS Patent 7,185,362.

\bibitem{gvili2020security}
Y.~Gvili, ``Security analysis of the covid-19 contact tracing specifications by
  apple inc. and google inc.'' \emph{IACR Cryptol. ePrint Arch.}, vol. 2020, p.
  428, 2020.

\bibitem{springer2020progressive}
A.~Springer and S.~Whittaker, ``Progressive disclosure: When, why, and how do
  users want algorithmic transparency information?'' \emph{ACM Transactions on
  Interactive Intelligent Systems (TiiS)}, vol.~10, no.~4, pp. 1--32, 2020.

\bibitem{kleinman2020digital}
R.~A. Kleinman and C.~Merkel, ``Digital contact tracing for covid-19,''
  \emph{CMAJ}, vol. 192, no.~24, pp. E653--E656, 2020.

\end{thebibliography}

\end{document}